\documentclass[journal]{IEEEtran}
\usepackage{graphicx}
\usepackage{subfigure}
\usepackage{float}
\usepackage{subfloat}
\usepackage{amsmath}
\usepackage{amssymb}
\usepackage{amsthm}
\usepackage{epstopdf}
\usepackage{cases}
\usepackage{color}
\usepackage{makecell}
\usepackage{cite}
\usepackage{algorithmic}
\usepackage[ruled]{algorithm2e}

\usepackage[normalem]{ulem}

\ifCLASSINFOpdf
\else
\fi
\usepackage{caption}
\usepackage{mathtools}

\begin{document}

%
\title{Joint Communication, Sensing and Computation enabled 6G Intelligent Machine System}
%
%
%

\author{Zhiyong Feng,~\IEEEmembership{Senior Member,~IEEE,}
	   Zhiqing Wei,~\IEEEmembership{Member,~IEEE,}
	   Xu Chen,~\IEEEmembership{Member,~IEEE,}
	   Heng Yang,~\IEEEmembership{Member,~IEEE,}
	   Qixun Zhang,~\IEEEmembership{Member,~IEEE,}
	   and Ping Zhang,~\IEEEmembership{Fellow,~IEEE}
	   
	   \thanks{This work was supported by the National Key Research and Development Program of China under Grants \{2020YFA0711300, 2020YFA0711303\} and the National Natural Science Foundation of China under Grants \{61941102\}.}
	   
	   \thanks{Z. Feng, Z. Wei, Xu Chen, Heng Yang and Qixun Zhang are with Beijing University of Posts and Telecommunications, Key Laboratory of Universal Wireless Communications, Ministry of Education, Beijing 100876, P. R. China (Email:\{fengzy, weizhiqing, chenxu96330, yangheng, zhangqixun\}@bupt.edu.cn).}
	   
	   \thanks{Ping Zhang is with Beijing University of Posts and Telecommunications, State Key Laboratory of Networking and Switching Technology, Beijing 100876, P. R. China (Email: pzhang@bupt.edu.cn).}
	   \thanks{Correponding author: {Zhiyong Feng, Xu Chen}}
	   
   	   }
       

%
%

\markboth{}%
{Shell \MakeLowercase{\textit{et al.}}: Bare Demo of IEEEtran.cls for IEEE Journals}
%



\maketitle

\pagestyle{empty}  
\thispagestyle{empty} 

\begin{abstract}
With the rapid development of the smart city, high-level autonomous driving, intelligent manufacturing, and etc., the stringent industrial-level requirements of the extremely low latency and high reliability for communication and new trends for sub-centimeter sensing have transcended the abilities of 5G and call for the development of 6G. Based on analyzing the  function design of the communication, sensing and the emerging intelligent computation systems, we propose the joint communication, sensing and computation (JCSC) framework for 6G intelligent machine-type communication (IMTC) network to realize low latency and high reliability of communication, highly accurate sensing and fast environment adaption. In the proposed JCSC framework, the communication, sensing and computation abilities cooperate to benefit each other by utilizing the unified hardware, resource and protocol design. Sensing information is exploited as priori information to enhance the reliability and latency performance of wireless communication and to optimize the resource utilization of the communication network, which further improves the distributed computation and cooperative sensing ability. We propose the promising enabling technologies such as joint communication and sensing (JCS) technique, JCSC wireless networking techniques and intelligent computation techniques. We also summarize the challenges to achieve the JCSC framework. Then, we introduce the intelligent flexible manufacturing as a typical use case of the IMTC with JCSC framework, where the enabling technologies are deployed. Finally, we present the simulation results to prove the feasibility of the JCSC framework by evaluating the JCS waveform, the JCSC enabled neighbor discovery (ND) and medium access control (MAC).

\end{abstract}

\begin{IEEEkeywords}
6G system, intelligent machine-type communication, joint communication, sensing and computation framework.
\end{IEEEkeywords}

%
\IEEEpeerreviewmaketitle


\begin{figure*}[!t]
	\centering
	\includegraphics[width=0.70\textheight]{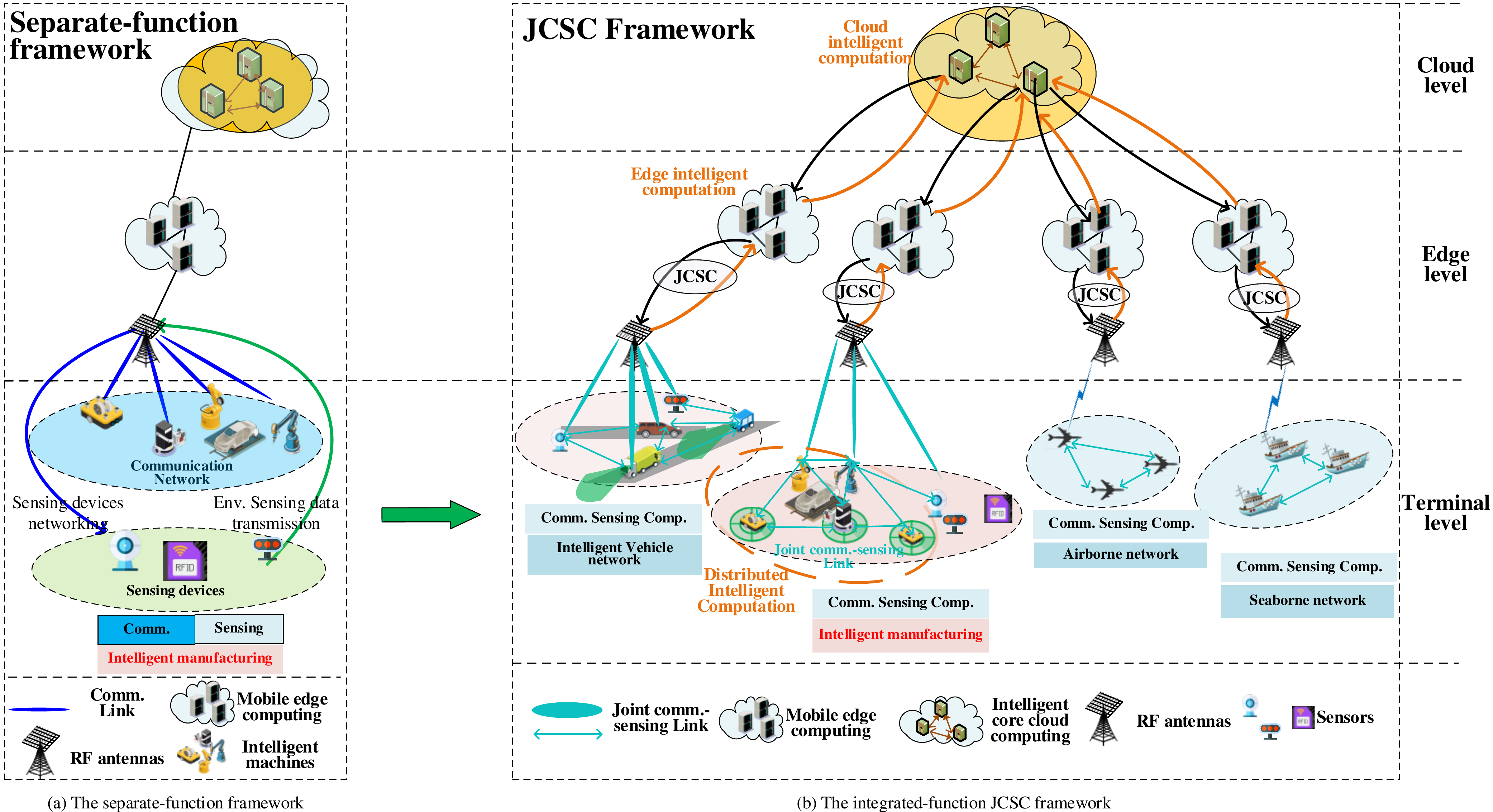}%
	\DeclareGraphicsExtensions.
	\caption{Two IMTC network frameworks: separate-function framework and integrated-function JCSC framework}
	\label{fig:JSC_framework}
\end{figure*}

\section{Introduction}
In order to satisfy the requirements on manufacturing, transportation and other fields of society, which are increasingly intelligent, autonomous, digitized and networked, the six generation (6G) mobile communication system intends to achieve ubiquitous wireless intelligence to support hyper-connected, digital and automatic intelligent applications in the future society~\cite{latva2019key}.  As one of the most challenging envisions, application scenarios such as smart transportation, smart city and smart industrial production have received unprecedented attention. In these scenarios, the intelligent machines (IMs) can autonomously learn from the experience knowledge and adapt to the environment by modifying the reaction strategies as the environment changes~\cite{InnoIM2007, christensen2016roadmap}. Hence, the intelligent-machine-type  communication (IMTC) will bring new growth motivation for the 6G mobile communication systems~\cite{mahmood2020six}.

In 6G era, IMs need to carry out complex precision tasks in highly dynamic environment. This requires IMs to achieve real-time environment sensing and autonomous cooperation with each other~\cite{mahmood2020white}. Thus, IMs need to be endowed with sensing and computing capabilities for environment sensing and decision-making to precisely and quickly form network with each other and conduct distributed collaborative computing~\cite{Chen2020PerCSUN}. Moreover, IMs are required to achieve close-loop control information flow including sensing data acquisition and sharing, intelligent processing and decision, deterministic control instruction transmission, and execution, etc.~\cite{You2021}. For example, in the scenario of  intelligent flexible manufacturing, the performance of close-loop control information flow has  a direct impact on the industrial level automatic control. This scenario requires extremely high reliability of more than 99.99999\%, and close-loop latency of 0.5 ms to 2 ms. The IMs also have the requirements of highly accurate sensing performance, extremely high uplink experience rate for the high-resolution sensing data transmission~\cite{latva2019key,Jiang2021RoadTo6G, Mourad2020, 2021Walid}. In this regard, the interactive messages among the neighbor machines and the uplink sensing data rate may reach several Gbps~\cite{Jiang2021RoadTo6G}. 

However, current mobile communication systems, represented by 5G system, are unable to support these accurate sensing, close-loop control information flow, and distributed collaborative computing  abilities. The main reasons are summarized as follows.
\begin{itemize}
		
	\item The communication, sensing and computation functions are separately designed to optimize each system individually, as shown in Fig.~\ref{fig:JSC_framework}(a), which will lead to the inconsistency among different functions, the non-negligible protocol translation overhead, large time-energy consumption in the interaction among these
	function modules. Hence, the performance optimization of close-loop control information flow is difficult to be guaranteed.

	\item The IMs are sensitive to the highly dynamic environment where multiple IMs need to accomplish critical tasks cooperatively. However, the tradition communication networks are not designed for sensing. The separated design of sensing and communication networks will degrade the performance of space-time-frequency resources scheduling, which does not support the high-efficient multiple nodes cooperative sensing.
	
	\item In the architecture of traditional communication network, the joint optimization of sensing and computation systems is not sufficient. The terminal IMs in the traditional architecture does not support the distributed networking and intelligent computing, which cannot meet the requirements of IMs for the autonomous cooperation and fast environment adaption.
	
\end{itemize} \par
Similar to biological intelligence, the intelligent close-loop control of IMs is supported by the interaction among environment sensing (sensory systems), communication (neural systems), and computational decision-making (brain). The joint design of communication, sensing and intelligent computation functions in the same framework, even the same equipment and protocols, can greatly improve the efficiency of resources utilization such as spectrum and energy, enable the resource sharing between communication and sensing, improve the throughput, latency and sensing accuracy according to the environment and task requirements, and reduce the information interaction overhead among these functions~\cite{liu2020joint,Chen2021CDOFDM, Fangzixi2020, 2021Young}. The communication network provides distributed intelligent computation abilities. The sensing ability provides massive environment data for intelligent network control and decision, which can enhance the efficiency of communication network reciprocally. All these features are beneficial to support the accurate sensing, close-loop control information flow and distributed collaborative computing requirements for 6G IMTC, which motivates the proposed joint communication, sensing and computation (JCSC) framework.

In this article, we first introduce the JCSC framework for 6G IMTC. Then, we discuss the enabling technologies and challenges for the realization of the JCSC framework. Subsequently, we give the evaluation results of the typical enabling techniques to show the effectiveness of JCSC framework. Finally, a brief conclusion and future work are summarized.

\section{JCSC Framework}

{\color{black}

Considering the challenges on the low latency and high reliability of close-loop
control information flow, the accurate sensing, and the collaborative computation, etc., for the 6G IMTC network, we propose the integrated communication, sensing and computation functions enabled JCSC framework, which is shown in Fig.~\ref{fig:JSC_framework}(b). Compared with the traditional separate-function framework, the sensing, communication and computation functions cooperation can improve the performance of close-loop control information flow, sensing performance and computing efficiency. Besides, the requirements on the uplink and downlink transmission are significantly different in capacity, latency and reliability, etc., because the uplink and downlink are mainly responsible for sensing data upload and instruction transmission, respectively. Thus, compared with the traditional separate-function framework, the JCSC framework can conduct the customized downlink and uplink transmission optimization to comply with various performance requirements. The novelty of the proposed JCSC framework is summarized as follows. 

\begin{itemize} 
	
	\item JCSC framework achieves the benefits among sensing, communication and computation functions. Sensing information can be exploited as priori information to enhance the performance of communication. The communication function supports the distributed intelligent computation and cooperative sensing ability of IMs. Finally, the intelligent computation can enhance the performance of sensing and communication.

	\item JCSC framework benefits the close-loop optimization of industrial application by regarding sensing, communication and intelligent information processing as a unified process in the JCSC enabled IMTC system, which satisfies the requirements of IMs for fast cooperation and environment adaption.

	\item A collaborative cloud-edge-terminal hierarchical decision	mechanism in the JCSC framework, i.e., global control at the cloud level, local control at the edge level, and autonomous decision making and sensing information acquisition at the terminal level, is proposed to support	the cloud intelligent computing, edge intelligent computing, collaborative sensing, and close-loop optimization.  The explanations for the cloud, edge and terminal levels are given as follows.
	
\end{itemize} \par

\emph{\it Terminal level}:
The terminal level consists of a large number of different types of IMs equipped with various sensory units (camera, radar, LiDAR, etc.). Besides, the joint communication and sensing (JCS) technique is utilized at the terminal level, which can use the same radio frequency (RF) transceiver in the same spectrum band to realize both communication and sensing functions,  simultaneously~\cite{liu2020joint,Zhang2021JCS5GCAV,Chen2021CDOFDM}. With the assisted sensing information of environments, such as the location of nodes acquired from JCS devices and sensing units, IMs can quickly conduct topology construction, medium access control (MAC), etc., to build the distributed or centralized self-organizing networks. Moreover, distributed intelligent computing, such as federal learning (FL), can also be conducted among IMs to achieve the massive sensing information integration for intelligent collaboration and network sensing, thus realizing the autonomous decision-making of IMs. Through distributed computing, JCSC enabled resource allocation and routing optimization can be achieved to further enhance the communication network performances.

\emph{\it Edge level}:
The edge level consists of base stations (BSs) and mobile edge computing servers (MECSs). The BSs utilize the JCS technique to conduct communication with IMs and sense the environment, simultaneously, while the MECSs interact with IMs at the terminal level via BSs. Besides, MECSs can receive the sensing information from different IMs within the coverage area of BSs to perform the regional sensing information integration and processing, and achieve the regional decision-making and control by utilizing machine learning methods, such as reinforcement learning (RL) and deep learning (DL). 

\emph{\it Cloud level}:
The cloud level consists of clustered servers to form the core cloud. The core cloud has the powerful computing ability and storage capacity. The core cloud servers can aggregate various data reported by IMs in the terminal level and MECSs in the edge level to integrate the information from different regions. Further, the core cloud servers can record and analyze the aggregated information from all regions to build a global management model library, realizing the global sensing, decision-making and task scheduling.

}

In summary, as the key information to realize 6G applications, the sensing information and the communication data are carried by using the unified protocol. In this way, the protocol interaction overhead between communication and sensing is greatly reduced. The location and motion state of the IMs can provide crucial priori information about the channel states for communication system, benefiting the beam alignment, channel estimation accuracy, medium access control, etc., to improve the communication reliability, optimizing the communication system efficiently using the AI algorithms deployed at the MECSs or the core cloud servers. Besides, the IMs' location and motion states can also reduce the feasible domain of the computation system for decision-making. The networking ability achieved by the communication function supports the cooperative sensing and computation. The cooperative sensing ability enhanced by AI methods can improve the sensing accuracy. Thus, JCSC framework achieves the mutual benefits among sensing, communication and computation abilities.

According to the above promising benefits, the JCS technique, rapid and reliable JCSC networking, and intelligent computation enhanced sensing and communication are considered as three significant enabling technologies, which will be introduced in the next section.

\section{Enabling Technologies}
In this section, we introduce the key enabling technologies mentioned above and the challenges on the way, which are promising to be the cornerstone of the JCSC framework in 6G era. From the perspective of the physical signal processing, the JCS technique is the key enabling technology, which is demonstrated in Section \ref{sec:JCS}. Based on the JCS technique that can provide significant priori sensing information for communication networking, the JCSC networking techniques are proposed in the Section \ref{sec:JCSC_networking}. In terms of achieving the aforementioned accurate sensing, close-loop control information flow in the complex IMTC scenarios, intelligent collaborative computation is necessary and can be well supported by the JCS technique and the JCSC networking techniques. Moreover, the intelligent computation can reciprocally enhance the performance of JCS and JCSC networking techniques by providing intelligent optimization and decision-making abilities, as shown in the Section \ref{sec:intelligent_computation}.

\subsection{Joint Communication and Sensing Technique}\label{sec:JCS}
The state-of-the-art JCS technique utilizes the same RF transceivers to realize wireless communication and sensing functions simultaneously, which can be divided into two classes. 

\begin{itemize}
	\item[$\bullet$] Sensing and communication functions are achieved physically in one system, but with separate spectrum, time slot or spatial resource.
	
	\item[$\bullet$] Sensing and communication functions are firmly integrated in one system by sharing the same hardware, spectrum, time slot, spatial resource and waveform, which utilize the existed reflection of communication signals to achieve additional sensing capability. 
\end{itemize}

The first class is the co-existed design of communication and sensing functions with separate resources to avoid the mutual interference. The second class is the focus of this article, which solidly improves the energy and spectrum utilization efficiencies. Then, we introduce the key techniques and challenges of the JCS technique.

\subsubsection{JCS Adaptive Waveform Design}
The traditional JCS waveform design mainly follows two paradigms: 

\begin{itemize}
	\item[$\bullet$] Sensing-centric waveform design: a small amount of communication data is embedded in the sensing waveform to maximize the sensing performance with low communication performance~\cite{liu2020joint}.
	
	\item[$\bullet$] Communication-centric waveform design: additional sensing ability is achieved by utilizing communication waveform to keep the intact communication performance with weakened sensing accuracy by random transmit signals~\cite{Chen2021CDOFDM}.
\end{itemize}

However, in the practical application scenarios of 6G, the performance requirements on the IMs' communication and sensing abilities vary with the tasks. Therefore, the above two traditional JCS waveform design schemes can not meet the requirements of the highly dynamic environment. {\color{black} A promising solution is to design an adaptive waveform that can jointly adjust communication and sensing performances. According to the real-time wireless environment and task requirements, the system can use machine learning methods, such as deep reinforcement learning (DRL), to make real-time decisions to adjust the parameters of the adaptive waveform, such as the channel coding, direct sequence spread spectrum code, physical frame, and etc, to adjust the communication and sensing performance. } 

\begin{figure}[!t]
	\centering
	\includegraphics[width=0.33\textheight]{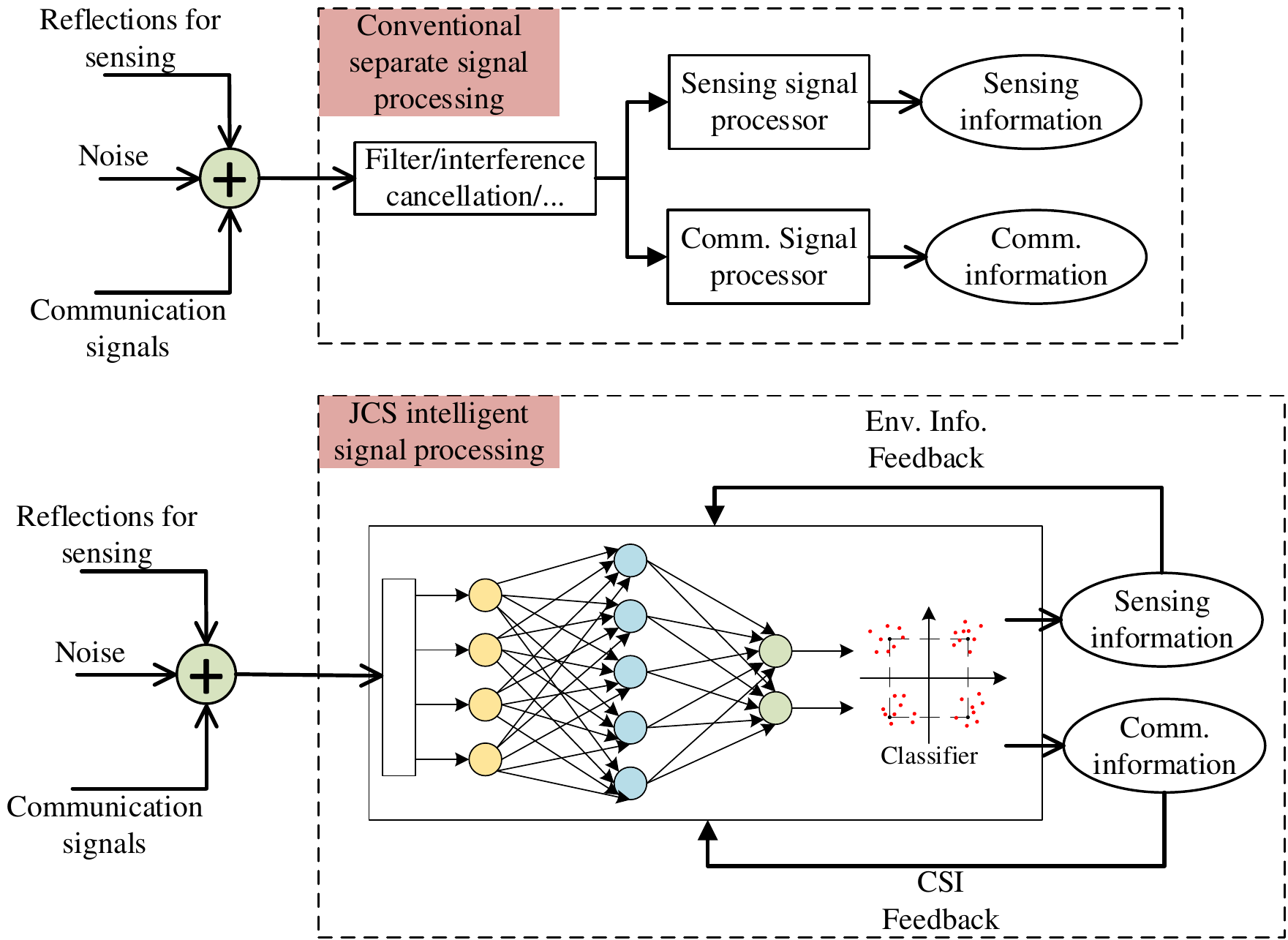}%
	\DeclareGraphicsExtensions.
	\caption{JCSC Intelligent signal processing technique}
	\label{fig:JSC_intelligent_procesing}
\end{figure}

\subsubsection{JCS Intelligent Signal Processing}

The processing of wireless communication and sensing signals is both the cascade of linear and nonlinear transform. In particular, the widely used orthogonal frequency division multiplex (OFDM) communication, frequency modulation continuous waveform (FMCW) and the OFDM radar signal processing all need to adopt fast Fourier transform (FFT)/ inverse FFT (IFFT) and classification methods to extract the communication and sensing information. This indicates that wireless communication and sensing information can be obtained simultaneously by using the data-driven AI methods to perform linear transform and intelligent non-linear signal classification for the JCS received signals. The comparison between the conventional JCS signal processing and the intelligent signal processing is shown in Fig.~\ref{fig:JSC_intelligent_procesing}. At the same time, it is highly possible that the correlation between communication CSI and sensing information, such as the range and Doppler, can be revealed. The realization of this technology will greatly reduce the computation load and processing delay of gaining the communication and sensing information.

\subsubsection{Challenges for JCS Technique}

The JCS technique faces multiple challenges in hardware implementation, signal processing, information theory, JCS performance metrics, and so on.
{\color{black} 
	\begin{itemize}
		
		\item[$\bullet$] \textit{Interference mitigation}: due to the high density and complex functions of equipment, the spectrum is congested, and the interference among JCS devices is intensified. It is urgent to study the intelligent interference mitigation techniques. 
		
		\item[$\bullet$] \textit{JCS adaptive waveform in highly dynamic environment}: the highly dynamic environment makes the requirements of IMs change rapidly. How to adjust the characteristics of the JCS adaptive waveform technique fast with a high performance is a key challenge. To solve this problem, the sensing information accuracy and refreshing rate should be high enough to provide better priori information.
		
		\item[$\bullet$] \textit{JCSC close-loop information theory}: in the field of information theory, the performances of point-to-point communication and radar sensing are studied respectively based on the traditional information theory. However, the 6G IMTC network needs to operate the close-loop process including sensing information transmission, intelligent information flow processing and action instruction transmission. It is urgent to explore the JCS close-loop information flow theory as well as the performance metrics for JCS, which can guide the close-loop optimization. 
		
	\end{itemize}
}

\subsection{JCSC Wireless Networking Techniques}\label{sec:JCSC_networking}

The complex and precision tasks require the IMs to cooperate with the specified IMs to conduct distributed intelligent computation efficiently, which requires the fast and precise networking ability with particular IMs. In the real tasks, the spatial location of each IM is the unique label to be differentiated from each other, which can be obtained by the sensing function of JCSC framework. Therefore, IMs are sensitive to the location and motion states of the neighbor IMs. With the sensing information as priori information, the JCSC framework improves the performance of neighbor discovery (ND), MAC, routing and resource allocation methods. Hence, JCSC technique endows IMs with environment awareness to assist fast and efficient cooperation among IM nodes, which can enable the highly efficient wireless, computation and hardware resources exploitation.

\subsubsection{JCSC Enabled Topology Construction}

The application of JCS technique endows the handshake frames of ND with sensing ability. Thus, the position of neighbors in the beam direction of an IM can be obtained. The discovery process of the potential neighbor IMs can be accelerated with the assistance of the sensing information collected by the edge IMs, which reduces the handshaking overload and network delay of ND among the IM network. With reinforcement learning methods, the IM adjusts the possibility of choosing a certain detection direction based on the sensing results in each direction, which improves the probability of successful ND.

\subsubsection{JCSC Enabled Medium Access Control}

The wireless and physical environment sensing information, obtained by the IMs and the edge server equipped with JCS equipment, assists the IMs to select the available spectrum and spatial resource blocks more efficiently, increasing the channel successful access rate. The sensing and communication functions share the same resources. To cope with the dynamic real-time requirements on these two functions, DRL method can be used to control the wireless access of JCS operation of different sorts of IMs through the intelligent control unit (CU) deployed in the edge cloud. As the iterative optimization of MAC policy continues, the JCSC MAC method can effectively reduce the collision probability and networking latency of the edge IMTC network, reduce the IMTC network congestion probability, and improve the reliability of communication links.

\subsubsection{JCSC Enabled Routing Method}

{\color{black}

The environment sensing information such as the location, motion state and network node density obtained by JCS techniques in the ND process provides significant priori information for the selection of relay nodes as well as the dynamic adjustment of the sensing information refreshing rate. This can   solve the broadcast storm problem caused by blind routing significantly and overcome the high relay latency problem caused by signal collision and congestion. 
} 

\subsubsection{JCSC Enabled Resource Allocation Method}
{\color{black} CUs in the JCSC framework utilize the priori information provided by the JCS equipment to learn the historical motion trajectory and spectrum usage of IM nodes, which could predict the communication service change, spectrum occupancy and the access probability of the IM nodes in the future. According to the prediction results, CUs can pre-allocate resources (e.g., transmit power, spectra) for IM nodes efficiently to reduce the access delay. In addition, due to the complexity of the environment, IM nodes running for a long time may break down at a certain moment. JCSC network can predict the possible failures of IMs based on the node running state information using supervised learning techniques. Then, the CUs pre-allocate resources for the backup IM nodes to replace the failure nodes, improving the robustness of the entire network.}

\subsubsection{Challenges for JCSC Wireless Networking Techniques}

\begin{itemize}
	
	\item The frame structure of the ND handshake-frame can affect the cross-correlation ranging performance of the sensing. Thus, the optimization of the design of the ND frame used to conduct JCS operation is a challenge, which can improve the positioning accuracy. 
	
	\item The IMTC networks in 6G era will be heterogeneous in functions, wireless resources, access methods, etc., and the network topology changes rapidly. Thus, to design the efficient JCSC MAC frame and intelligent random access mechanism in the high heterogeneity and dynamic environment is a great challenge.
	
	\item {\color{black}
		The JCSC network may cause a new problem in the routing process. The MAC address and the sensing data of the same IM node is hard to identify, which draws a challenge to the confirmation of the relay node. The association between the MAC address and the sensing data effectively is another challenge.}
	
	\item As the available spectrum for the JCSC network becomes much wider, the channel propagation properties on different frequency bands are significantly different. How to allocate the spectrum resources intelligently for sensing and communication functions according to the environment conditions is another challenging problem.
	
	\item JCSC network is a complex heterogeneous network with both centralized and distributed organizations. Hence, it is a challenge on the low-complexity optimization algorithm design to achieve the high-performance sensing and the low-latency communication with a low energy consumption.
	
\end{itemize} \par

\begin{figure*}[!t]
	\centering
	\includegraphics[width=0.58\textheight]{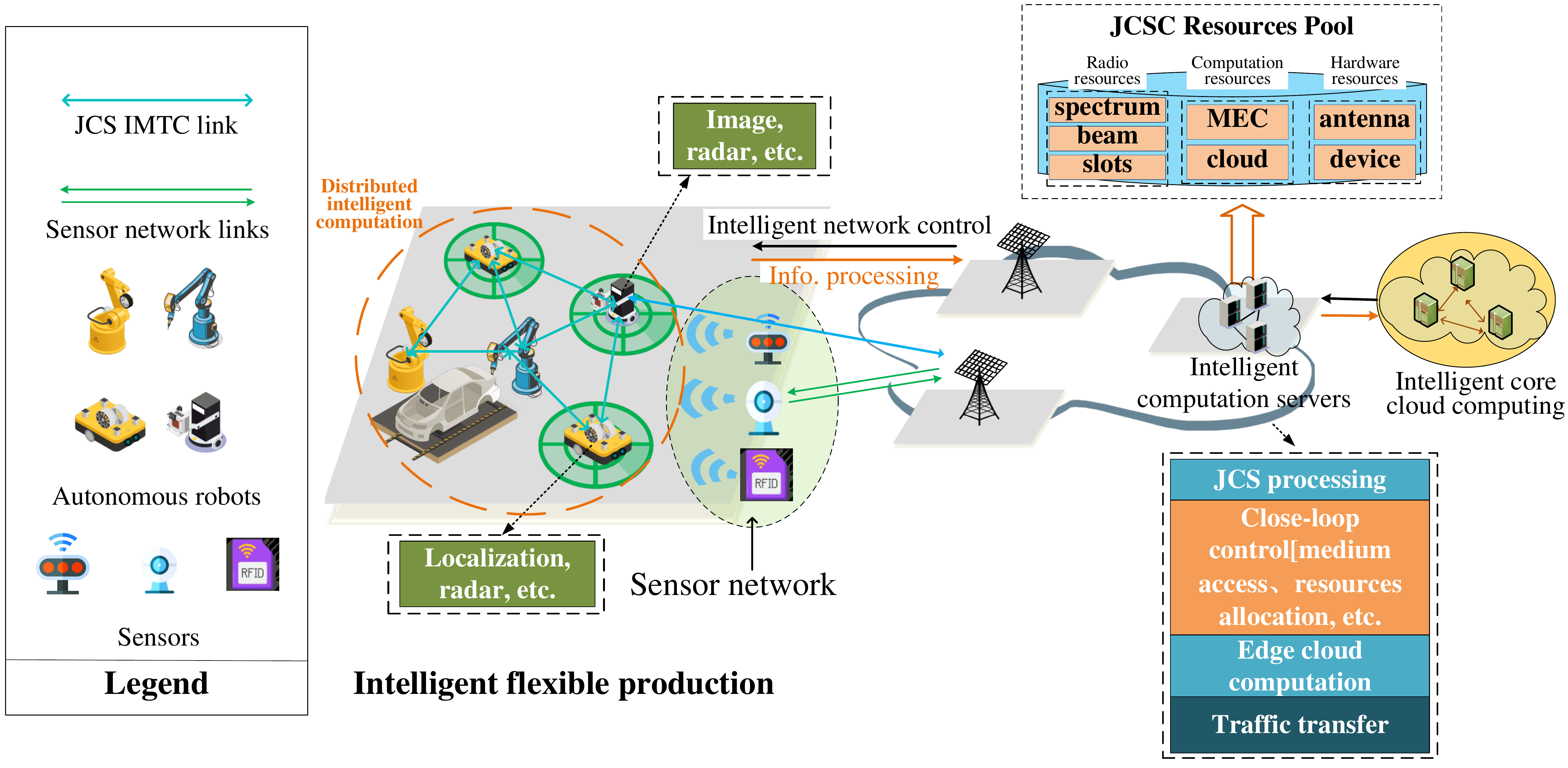}
	\DeclareGraphicsExtensions.
	\caption{JCSC intelligent flexible production scenario. The JCS technique is implemented in the base stations as well as in the IMs. The sensors are deployed to form a sensor network. Distributed computation and sensing are conducted among IMs. The edge and core clouds integrate the sensing information from IMs and sensors, and control the network.}
	\label{fig:Intelligent Production}
\end{figure*}

\subsection{Intelligent Computation Enhanced Communication and Sensing Techniques}\label{sec:intelligent_computation}

In the proposed JCSC network framework, 
the distributed and centralized intelligent computing can perform sensing data fusion and matching to achieve a better sensing vision and more accurate environment sensing information. The sensing ability and the communication resource pre-allocation ability can be enhanced thereafter. In this subsection, we introduce the key techniques and challenges of the JCSC computing technology. 

\subsubsection{Intelligent Computation Enhanced Communication-assisted Sensing}

At the edge side of IMTC network, multiple IMs can share locally processed sensing information such as the location and the type of objects, which is derived from the multi-source raw sensing data, with each other or with the edge access points via JCS devices. These sensing information can be fused at the center IM or edge/center cloud servers. The FL framework can also be used to train distributed intelligent learning models. When the sensing data is fused, AI methods can be operated to recognize and localize the targets in the environment, aggregate the discrete points into the real objects for IM applications. In addition, the future motion states can be predicted based on the current sensing information. The problem of blind sensing area can be alleviated, and more priori knowledge can be gained to enhance the communication and computation functions. 

\subsubsection{Intelligent Computation Enhanced Sensing-assisted Communication}

Taking the environment sensing information such as position and velocity of nodes as priori information, AI methods, such as deep learning (DL), can be used to enhance channel estimation and prediction, which improves beamforming performance. Moreover, using reinforcement learning (RL) method to assist beam alignment will greatly enhance the communication reliability and power efficiency. In addition, sensing information, such as the position of nodes, is beneficial to predicting the signal to interference plus noise ratio (SINR) and can assist power allocation, beam-user association, etc.

\subsubsection{JCSC Cloud-Edge-Terminal Collaborative Hierarchical Decision}

In the JCSC enabled IMTC network, some IMs have certain sensing and computing capabilities, while the MEC servers at the edge side and the cloud side have powerful computing capabilities. Hence, the IMs pre-process its own sensing information and deliver the processing results to the edge side. The MEC servers can realize the multi-level sensing data fusion of various sensor data and utilize DRL methods to achieve local or global autonomous decision-making and send the control information to each IM. Then, IMs make real-time autonomous decision-making based on the control information from the edge side and the cloud side.
 
\subsubsection{Challenges for Intelligent Computation Enhanced Communication and Sensing Technology}

\begin{itemize}
	
	\item The environment of the IMs is highly dynamic, and the characteristics of channels among nodes are impacted by complex factors. It is challenging to find the best categories of environment sensing information and the optimal deep learning architecture to predict channel changes accurately.
	
	\item Before the intelligent learning models of the IMTC network are trained to convergence, there may be errors in the sensing information, which will lead to the poor initial cooperative sensing and unreliable prediction. How to accelerate the convergence of the intelligent learning models and to determine the availability of  prediction based on the reliability requirements of applications are challenging. 
	
	\item Due to the relatively weak computing power of the edge IMs, the key issue is how to define the amount of information to be pre-processed by the IMs based on the dynamic wireless channel conditions. Another issue is how to perform multi-level sensing data fusion efficiently, while ensuring a low close-loop decision-making delay and a high detection accuracy.
	
\end{itemize} \par

\section{Typical Use Case}
In this section, the intelligent flexible manufacturing as a typical use case of the IMTC with JCSC framework is introduced, which is shown in Fig.~\ref{fig:Intelligent Production}. 

\begin{figure*}[!ht]
	\centering
	\includegraphics[width=0.65\textheight]{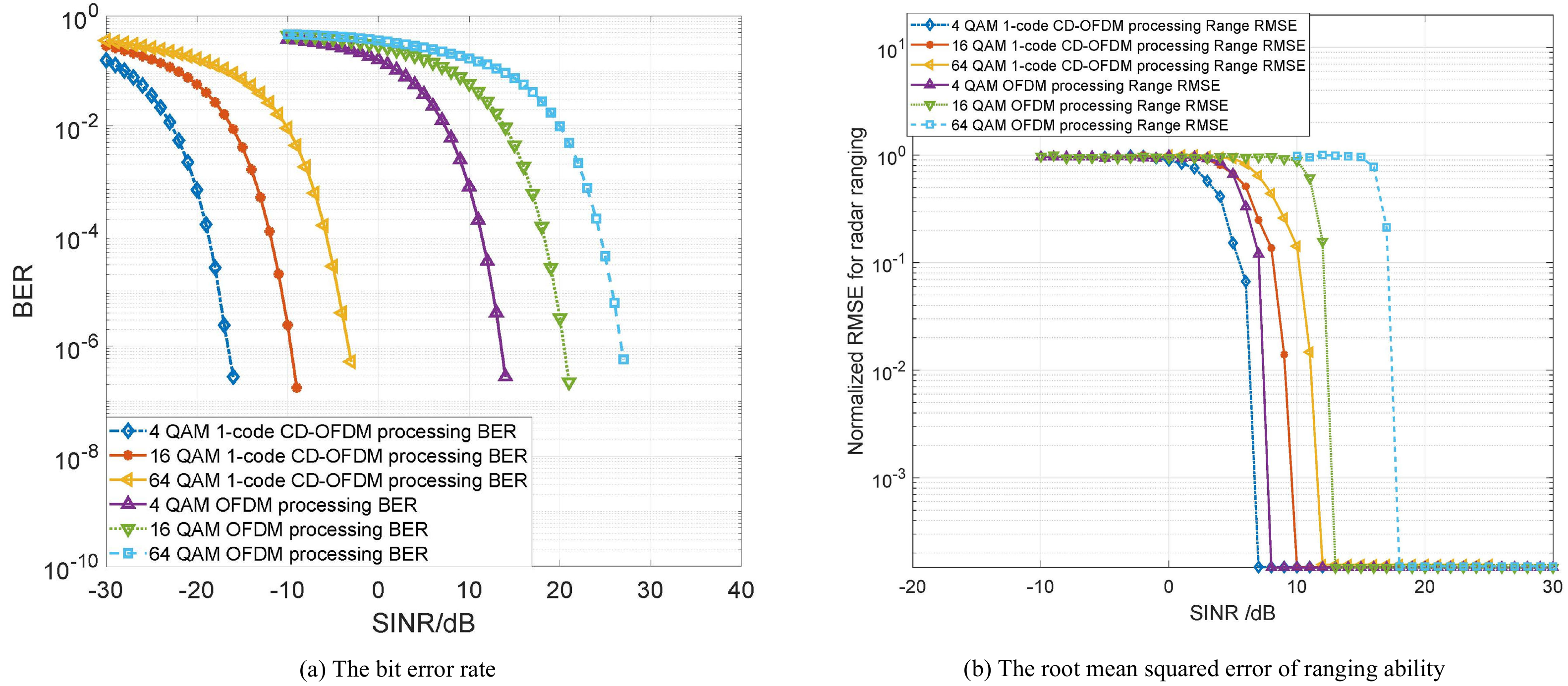}
	\DeclareGraphicsExtensions.
	\caption{Simulation results of the proposed CD-OFDM JCS waveform compared with the conventional OFDM JCS waveform~\cite{Chen2021CDOFDM}}
	\label{fig:JCS_technique}
\end{figure*}

In the scenario of intelligent flexible manufacturing, the IMs consist of autonomous vehicles, robots, etc. The highly efficient cooperation and environment adaption of IMs require extremely high reliability and low latency of the close-loop control information flow, high-accuracy sensing performance, and the efficient computing capability. All these requirements aim to achieve autonomous precision production through the cooperation of IMs and accelerate the speed of production.

In the JCSC framework, the JCS technique is implemented in the base stations as well as in the IMs to establish communication links among IMs and conduct wireless sensing over the environment simultaneously. The sensors such as cameras and RFIDs are deployed to form a sensor network according to the applications. The distributed sensing supported by the network among IMs is carried out to achieve fast and accurate sensing performance. The distributed intelligent computing is carried out to improve the autonomous decision-making ability of IMs. The sensing information obtained by IMs and the sensor network could be fused and processed at the edge computing servers to obtain comprehensive and accurate sensing results and to provide priori information for edge network control. The cloud intelligent computing unit is deployed for the macroscopic environment information processing and decision-making.

The radio resources, computation resources and hardware resources are virtualized and mapped into an integrated resource pool for the JCSC enabled IMTC. The radio resources are continuously sensed by the JCS units. Thus, the information of available spectrum, beams, slots, etc. is stored in the resource pool. In JCSC framework, the resources are jointly managed to reveal the optimal optimization of the close-loop information interaction, sensing and computation. The sensing information obtained by IMs is gathered at the edge server for the building of a centralized learning network. The sensing information can also provide the priori information for beam alignment, ND, MAC, routing, and network resource allocation, etc., which significantly improves the reliability, throughput and latency performance of communication. With the optimization of communication performance, the IMs will further improve the collaborative sensing and computing abilities, so that the efficiency of sensing, communication and computing will gradually improve interactively and satisfy the demand of intelligent flexible manufacturing in the era of 6G.

\section{Evaluation}

{\color{black}
In this section, we present the evaluation results of the novel JCS waveform, JCSC enabled topology construction, and JCSC enabled MAC.}

We apply code division multiplex (CDM) technique in the frequency domain symbols of OFDM waveform, namely, CD-OFDM, to achieve the JCS operation to obtain the CDM gain for both communication and sensing~\cite{Chen2021CDOFDM}. The simulation results of the bit error rate (BER) and root mean squared error (RMSE) of the proposed JCS system are shown in the Figs.~\ref{fig:BER_1_Noncode} and \ref{fig:RMSE_range_1_Noncode}. {\color{black} The proposed JCS waveform achieves better BER and RMSE than the conventional OFDM JCS waveform processing mainly because the CDM gain improves the reliability of communication demodulation and the ratio of echo signals to the interference plus noise}. In the simulation, with uniform linear array, the operating frequency is 24 GHz with 122.88 MHz bandwidth. Compared with the conventional OFDM JCS processing~\cite{Chen2021CDOFDM}, the proposed CD-OFDM JCS processing can achieve around 30 dB CDM gain in communication, and 1 to 6 dB gain in sensing performance. {\color{black} In the aspect of JCS waveform design, the code division mechanism enjoys the reliability gain at the cost of computation resources, it remains a problem to model the trade-off between the computation and JCS performance.} 

\begin{figure}[!t]
	\centering
	\includegraphics[width=0.30\textheight]{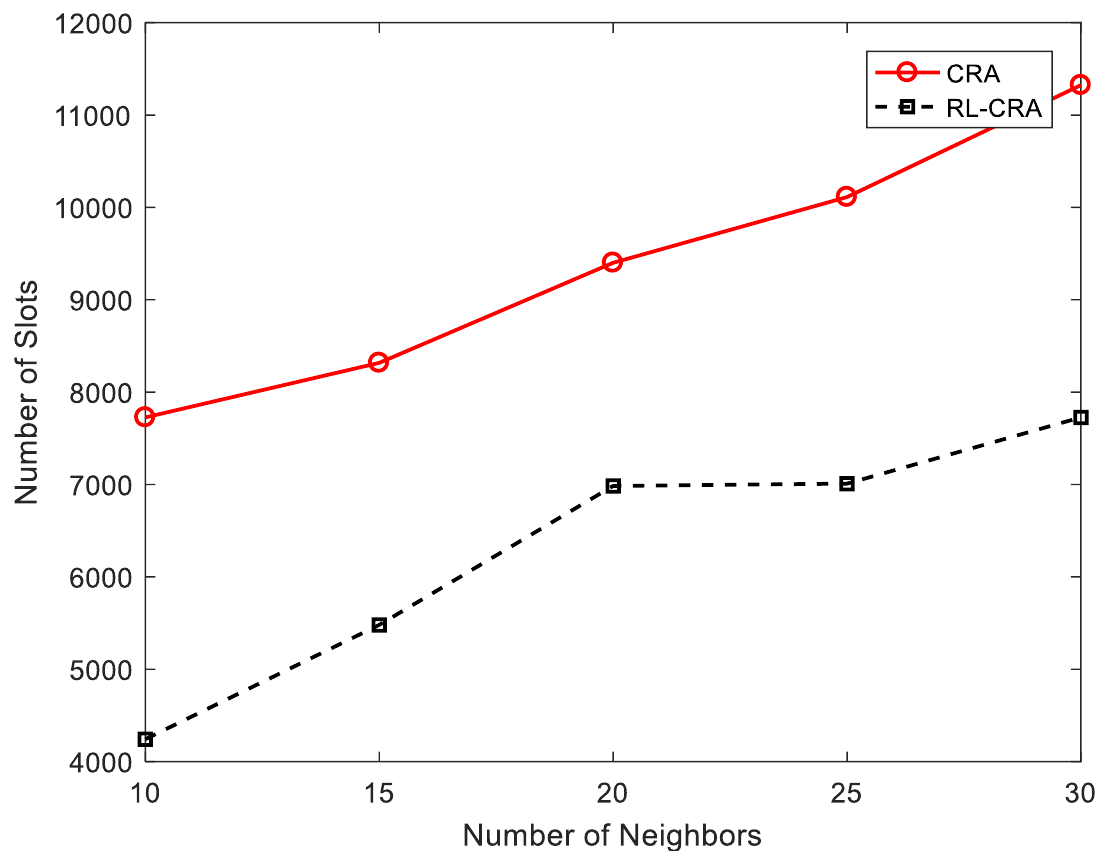}%
	\DeclareGraphicsExtensions.
	\caption{The time slots required by the conventional complete random scanning (CRA) method and the proposed RL-CRA method corresponding to the JCSC framework}
	\label{fig:ND_Performance}
\end{figure}

In the JCSC ND process, with the priori sensing information, the ND process can be accelerated. The performance comparison between the traditional complete random scanning (CRA) ND algorithm and the sensing-enhanced RL-based CRA (RL-CRA) algorithm is shown in Fig.~\ref{fig:ND_Performance}. Here, the ratio of the effective sensing distance of the radar sensing to the effective communication distance is 1/2, and the beamwidth of the JCS transceiver is 10 degrees. When the number of neighbor nodes is 30, compared with the traditional CRA method, the RL-CRA algorithm can reduce the ND process delay by 31.7\%.

\begin{figure}[!t]
	\centering
	\includegraphics[width=0.33\textheight]{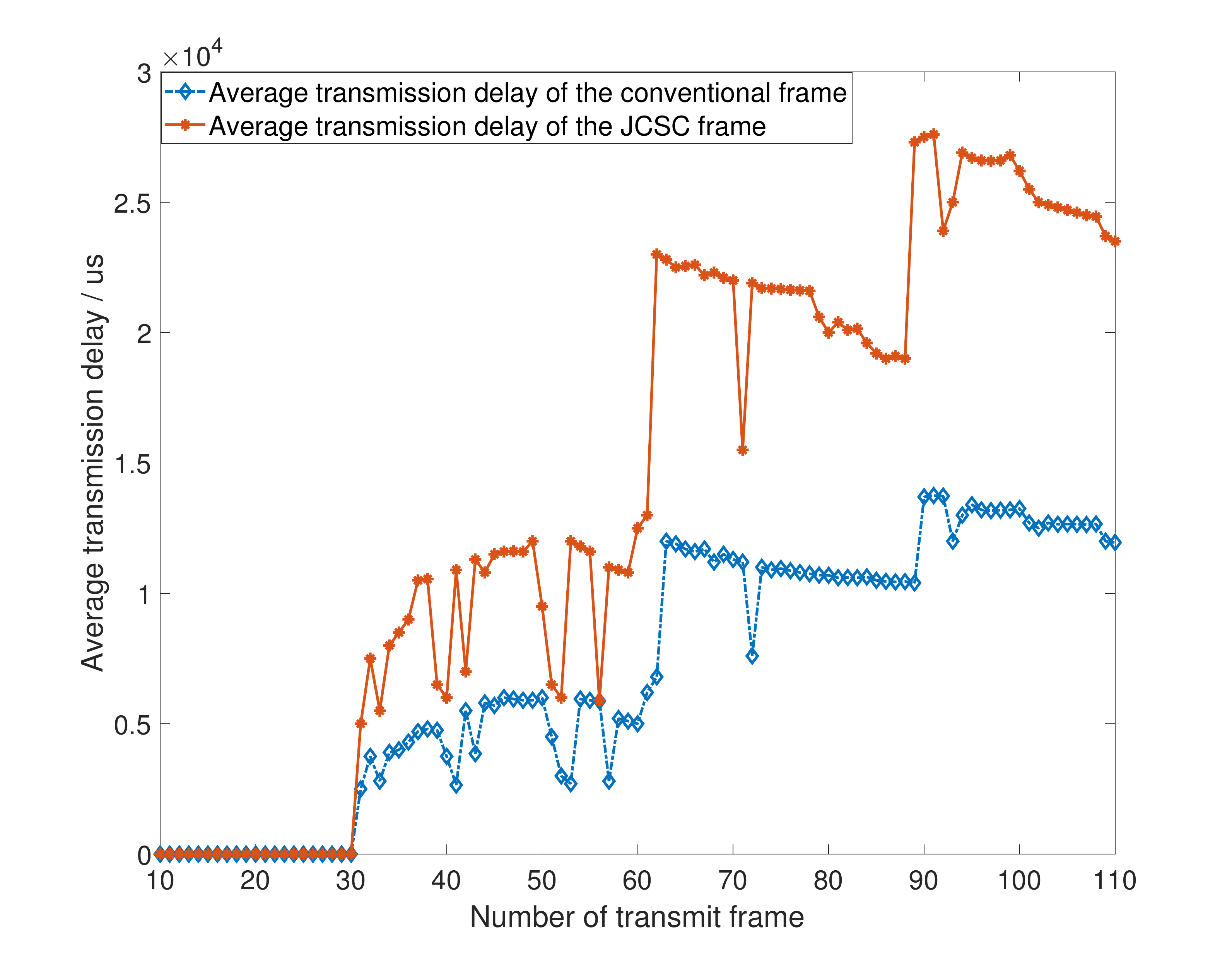}%
	\DeclareGraphicsExtensions.
	\caption{Data frame transmission latency}
	\label{fig:Data Frame Transmission latency}
\end{figure}

In the JCSC MAC process, the potential data collision can be avoided, because a certain number of hidden nodes can be detected by the above JCSC ND process. Thus, the data transmission delay can also be reduced. Fig.~\ref{fig:Data Frame Transmission latency} illustrates the average network data frame transmission delay varying with the time length of transmitted frames. The number of nodes in the network is 10. Compared with the conventional method shown by the blue curve, the JCSC network can reduce the average data transmission delay by around 50\%.  

\section{Conclusion}

In this paper, we have proposed a JCSC framework to satisfy the extremely low latency and high reliability requirements for 6G intelligent machine system. Potential enabling technologies are summarized and the challenging problems are also outlined. Typical use case of intelligent flexible manufacturing is introduced and numerical results verify the feasibility of the proposed JCSC technologies. 

This paper has opened a window for the innovations of 6G intelligent machine system. In the future, there are still many challenges for the implementation of JCSC. The relationship, the theoretical bound and the performance trade-off among communication, sensing and computing functions are the fundamental problems to be solved. The JCSC system should also move forward to a deep integration stage in terms of the physical, MAC, and network layers. In summary, the proposed JCSC framework is one of the key enabling technologies for 6G communication system, which has a wide application prospect in autonomous driving, intelligent manufacturing, intelligent robotics, and etc. 


\vspace{5 mm}
	{\textbf{Zhiyong Feng}} (M'08-SM'15) received her B.E., M.E., and Ph.D. degrees from Beijing University of Posts and Telecommunications (BUPT), Beijing, China. She is a senior member of IEEE, vice chair of the Information and Communication Test Committee of the Chinese Institute of Communications (CIC).
	
\vspace{0 mm}
	{\textbf{Zhiqing Wei}} (S'12-M'15) received his B.E. and Ph.D. degrees from Beijing University of Posts and Telecommunications (BUPT) in 2010 and 2015. He was granted the Exemplary Reviewer of IEEE Wireless Communications Letters in 2017, the Best Paper Award of International Conference on Wireless Communications and Signal Processing (WCSP) 2018.
	
\vspace{0 mm}
	{\textbf{Xu Chen}} (S'18) received the B.S. degree from Southwest Jiaotong University in 2018. He is now in pursuit of the Ph.D. degree from Beijing University of Posts and Telecommunications. His research interest includes wireless communication, machine-type communication, sensor network, joint communication and sensing network, and signal processing.
	
\vspace{0 mm}
	{\textbf{Heng Yang}} (S'20) received the B.Sc. degree and the M.Sc. degree from Chongqing University of Posts and Telecommunications, Chongqing, China, in 2016 and 2019. He is currently pursuing the Ph.D. degree with the School of Information and Communication Engineering, Beijing University of Posts and Telecommunications, Beijing, China. 

	{\textbf{Qixun Zhang}} (M'12) received the B.Eng. degree in communication engineering and the Ph.D. degree in circuit and system from Beijing University of Posts and Telecommunications (BUPT), China, in 2006 and 2011, respectively. From Mar. to Jun. 2006, he was a Visiting Scholar at the University of Maryland, College Park, Maryland. 

	{\textbf{Ping Zhang}} (Fellow, IEEE) received the M.S. degree in electrical engineering from Northwestern Polytechnical University, Xi’an, China, in 1986, and the Ph.D. degree in electric circuits and systems from the Beijing University of Posts and Telecommunications (BUPT), Beijing, China, in 1990. He is currently a Professor with BUPT.

%

%
%

\ifCLASSOPTIONcaptionsoff
  \newpage
\fi

\end{document}